\documentclass[aps,showpacs,bibnotes,twocolumn]{revtex4}
\usepackage{graphicx}
\begin{document}
\draft
\def\ds{\displaystyle}
\title{Critical Trap Aspect Ratios For Dipolar BEC}
\author{C. Yuce}
\address{ Physics Department, Anadolu University,
 Eskisehir, Turkey}
\email{cyuce@anadolu.edu.tr}
\date{\today}
\begin{abstract}
We show that there exists critical trap aspect ratios for a
trapped Bose-Einstein condensate with dipole-dipole interactions.
We discuss the role of critical trap aspect ratios on both the
critical angular velocity above which a vortex is energetically
favorable and the precession velocity of an off-axis vortex in TF
regime. We show that the stability diagram for a purely dipolar
gas depends crucially on the critical trap aspect ratios both for
positive and negative strength of dipolar interaction. We prove
that the critical points can be changed by an anharmonic
potential.
\end{abstract}\maketitle

\section{Introduction}

The ultracold gases with dipole-dipole interaction became the
subject of intensive theoretical studies after the first gaseous
condensate was produced. The experimental realization of
Bose-Einstein condensation of $\ds{^{52}}$Cr atoms in 2005 has
raised the interest in study of BEC with nonlocal dipole-dipole
interactions \cite{baro1,baro2,baro3,baro4}. This nonlocal
character of the potential has remarkable consequences for the
physics of rotating dipolar gases in TF limit. It is believed
that, in axially symmetric traps with the axis along the dipole
orientation, the critical angular velocity, above which a vortex
is energetically favorable, is decreased due to the dipolar
interaction in oblate traps, and increased in prolate traps
\cite{TFvortex,rev1,baranov}. It was noted in \cite{cem} that the
effect of the dipole-dipole interaction is the lowered (raised)
precession velocity of an off-center straight vortex line
in an oblate (a prolate) trap. \\
The nonlocal character of dipolar potential is also crucial for
the stability properties of dipolar gases. It was shown that there
exists a critical aspect ratio in the stability
diagram of a purely dipolar condensate \cite{gen8}.\\
In this paper, we show that the crossover between a reduced and an
increased critical angular velocity above which a vortex is
energetically favorable occurs at a critical aspect ratio in TF
regime. We discuss the role of the critical aspect ratio on the
precession velocity of an off-center straight vortex in TF limit.
In the Gaussian limit, we show that there exists two critical
aspect ratios for a purely dipolar condensate. We investigate
their role on the stability diagram. We discuss that an additional
anharmonic potential changes the
critical trap aspect ratios. \\
The paper is organized as follows: The following section reviews
two main methods to calculate the dipolar potential interaction.
Section III finds a critical trap aspect ratio in TF limit.
Section IV investigates critical trap aspect ratio in the
stability diagram and discusses the effect of additional quartic
potential on the critical aspect ratios.\\

\section{Dipolar Interaction}

Consider a BEC of N particles with mass $\ds{m}$ and magnetic
dipole moment $\ds{\mu}$ oriented in the z-direction by a
sufficiently large external field. At sufficiently low
temperatures, the description of the ground state of the
condensate is provided by the solution of the Gross-Pitaevskii
(GP) equation
\begin{equation}\label{denklem1}
\left(-\frac{\hbar^2}{2m} \nabla^2 +V_{T}+g |\Psi|^2+\Phi_{dd}
\right)\Psi=\mu\Psi~,
\end{equation}
where $\ds{\Phi_{dd}(\textbf{r})={\frac{C_{dd}}{4 \pi}} \int
d^3\mathbf{r^{\prime}} \frac{1-3 \cos^2
\theta}{|\mathbf{r}-\mathbf{r^{\prime}}|^3}
|\Psi(\mathbf{r^{\prime}},t)|^2}$ is the mean field term due to
dipole-dipole interactions, $\ds{C_{dd}}$ is the strength of the
dipolar interactions, $\ds{\theta}$ is the angle between the
vector connecting two dipolar particles and the dipole
orientation, $\ds{g=\frac{4\pi \hbar^2a_s}{m}}$, $\ds{a_s}$ is the
scattering length, $\ds{V_{T}}$ is the trap potential
\begin{equation}\label{vrd55}
V_{T}(\mathbf{r})=\frac{m}{2}\omega_{\perp}^2\left(r^2+
~{\gamma^2} z^2\right)~,
\end{equation}
where $\ds{\gamma=\frac{\omega_{z}}{\omega_{\perp}}}$ is the trap
aspect ratio. It is well known that the sign of the dipolar
mean-field energy can be controlled via the trap aspect ratio. For
dipolar condensates, it is useful to introduce a dimensionless
parameter that measures the relative strength of the dipolar and
s-wave interactions $\ds{\varepsilon_{dd}=\frac{C_{dd}}{3 g } }$.
Chromium atoms posses an anomalously large magnetic dipole moment,
hence $\ds{\varepsilon=0.16 }$. It can be enhanced via Feshbach
tuning of the scattering length. In Thomas-Fermi regime, dipolar
BEC is stable as long as $\ds{-0.5<\varepsilon_{dd}<1}$.\\
The equation (\ref{denklem1}) is an integro-differential equation
since it has both integrals and derivatives of an unknown wave
function. Let us focus on the dipolar potential term,
$\ds{\Phi_{dd}}$. There are two main methods to calculate
$\ds{\Phi_{dd}}$. The first one is based on a Gaussian trial
function. $\ds{\Phi_{dd}}$ can be calculated by employing Fourier
transform from $\ds{(r,z)}$-space to
$\ds{(k_{\perp},k_{z})}$-space. The Fourier transform of the
dipole-dipole interaction term in the small cut-off distance is
given by $\mathcal{F}\left\{\frac{1-3 \cos^2
\theta}{|\mathbf{r}-\mathbf{r^{\prime}}|^3}\right\}=\frac{4
\pi}{3}~ (3 \cos^2\alpha-1)$  \cite{goral,ftpot}. Hence, the
convolution theorem leads
\begin{equation}\label{dfgk434598}
\Phi_{dd}=\frac{C_{dd}}{3}\mathcal{F}^{-1}\left\{ (3
\cos^2\alpha-1) \mathcal{F}\left\{|\Psi|^2\right\}\right\}
\end{equation}
where $\ds{\mathcal{F}}$ and $\ds{\mathcal{F}^{-1}}$ denote the
Fourier and the inverse Fourier transforms, respectively and
$\ds{\alpha}$ is the angle between the momentum $\ds{\textbf{k}}$
and the dipole direction.  Since we have assumed that the dipoles
are polarized along the z-axis,
$\ds{\cos^2\alpha=\frac{k_z^2}{k_z^2+k_{\perp}^2}}$. The
corresponding dipole-dipole energy can be calculated using eq.
(\ref{dfgk434598}) and the Parseval's theorem
\begin{equation}\label{d523}
E_{dd}=\frac{{C_{dd}}}{6} \int (3 \cos^2\alpha-1)
\left|\mathcal{F}\{|\Psi|^2\}\right|^2 d^3\textbf{k}
\end{equation}
where $\ds{d^3\textbf{k}=2 \pi ~k_{\perp}dk_{\perp}~dk_z }$. The
last relation is of importance in the variational method, which is
an efficient method to study the stability properties and
collective excitations of dipolar gas
\cite{gen8,goral,ftpot,gen1,gen3,gen7,stability1,stability2,stability3,ek1,ek2}.\\
The second method to calculate $\ds{\Phi_{dd}}$ was presented by
Eberlein, Giovanazzi and O'Dell in TF regime \cite{TF}. They
showed that a parabolic density profile remains an exact solution
for an harmonically trapped vortex-free dipolar condensate in the
TF limit. In the TF regime, the mean-field dipolar potential,
$\ds{\Phi_{dd}(\textbf{r})}$, is given by
\begin{equation}\label{fidd}
\Phi_{dd}(\rho,z)=\frac{n_{0} C_{dd}}{3} \left(
\frac{\rho^{2}}{{R}^{2}}-\frac{2
z^{2}}{{L}^{2}}-f(\kappa)(1-\frac{3}{2} \frac{\rho^{2}-2
z^{2}}{{R}^{2}-{L}^{2}})   \right)
\end{equation}
where the function $\ds{f(\kappa)}$ for oblate case, ($\kappa>1$),
is given by $\ds{f(\kappa)=\frac{2+\kappa^{2}(4-6~
\frac{\arctan{\sqrt{\kappa^{2}-1}}}{\sqrt{\kappa^{2}-1}}~)}{2
(1-\kappa^{2})}}$. Here, $\ds{\kappa=\frac{R}{L}}$ is the
condensate aspect ratio. In the absence of dipolar interaction,
the condensate aspect ratio and the trap aspect ratio match.

\section{A Single Vortex}

\begin{figure}[htp]
\includegraphics[width=6cm]{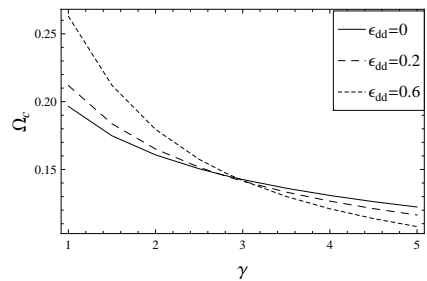}
\caption{\label{fig:CriticalVelocity} The critical angular
velocity of a condensate for $\ds{\varepsilon_{dd}=0}$,
$\ds{\varepsilon_{dd}=0.2}$, and $\ds{\varepsilon_{dd}=0.6}$ as a
function of trap aspect ratio $\ds{\gamma}$. $\ds{\Omega_{c}}$ is
measured in units of $\ds{\omega_{\perp}}$. $\ds{\Omega_{c}}$
increases (decreases) with $\ds{\varepsilon_{dd}}$ for
$\ds{\gamma<\gamma_c}$ ($\ds{\gamma>\gamma_c}$), where
$\ds{\gamma_c\approx2.8}$}
\end{figure}

The density profile of a dipolar condensate with a straight
central vortex line in the TF regime reads \cite{TFvortex}
\begin{equation}\label{density}
n(\textbf{r})=n_{0}\frac{ \rho^2}{ \rho
^2+\beta^{2}}\left(1-\frac{\rho^{2}}{R^{2}}-\frac{z^{2}}{L^{2}}\right)
\end{equation}
where $\ds{n(\textbf{r})=0}$ when the right hand side is negative
and $\ds{\beta}$, $\ds{R}$, and $\ds{L}$ are variational
parameters that describe the size of the vortex core, the radial
and the axial sizes, respectively. O'Dell and Eberlein derived the
corresponding energy expression elegantly and performed numerical
calculations for a prolate trap with $\ds{\gamma=0.2}$ and an
oblate trap with $\ds{\gamma=5}$ \cite{TFvortex}. They found that
the critical angular velocity, $\ds{\Omega_c}$, above which a
vortex state is energetically favorable, is increased for the
former one and decreased for the latter one. Based on the results
for $\ds{\gamma=0.2}$ and $\ds{\gamma=5}$, the authors concluded
that $\ds{\Omega_c}$ is decreased due to the dipole-dipole
interaction in oblate traps, $\ds{\gamma>1}$, and increased in
prolate traps, $\ds{\gamma<1}$. We find that this generalization
does not hold. The crossover between a reduced and an increased
critical velocity occurs at $\ds{\gamma=\gamma_c>1}$ rather than
$\gamma=1$ in TF regime. In other words, $\ds{\Omega_c}$ are the
same for dipolar and non-dipolar condensates at $\ds{\gamma_c}$.
The critical value $\ds{\gamma_c}$ changes slightly with
$\epsilon_{dd}$. For $\ds{^{52}}$Cr atoms with
$\ds{\epsilon_{dd}=0.16}$, $\ds{\gamma_c{\approx}2.8}$. For
$\ds{\varepsilon_{dd}=0.6}$, $\ds{\gamma_c\approx2.9}$.
Fig-\ref{fig:CriticalVelocity} plots $\ds{\Omega_{c}}$ in units of
$\ds{\omega_{\perp}}$ as a function of trap aspect ratio
$\ds{\gamma}$ for three different values of
$\ds{\varepsilon_{dd}}$ ($\ds{\varepsilon_{dd}=0}$,
$\ds{\varepsilon_{dd}=0.2}$, and $\ds{\varepsilon_{dd}=0.6}$). As
can be seen from the figure, $\ds{\Omega_{c}}$ increases with
$\ds{\varepsilon_{dd}}$ for $\ds{\gamma<\gamma_c}$, while
$\ds{\Omega_{c}}$ decreases with $\ds{\varepsilon_{dd}}$ for
$\ds{\gamma>\gamma_c}$.\\
So far, we have focused on a central vortex. In \cite{cem}, based
on two examples, it was concluded that the effect of the
dipole-dipole interaction is the lowered (raised) precession
velocity of an off-center straight vortex line in an oblate (a
prolate) trap. We conclude that the effect of the dipole-dipole
interaction is the lowered (raised) precession velocity of an
off-center straight vortex line
for $\gamma<\gamma_c$ ($\gamma>\gamma_c$).\\
Having found a critical trap aspect ratio in TF limit, let us
search for the existence of a critical trap aspect ratio in
Gaussian approximation.

\section{Stability}

In this section, we will explore another critical value missed in
earlier papers investigating stability properties of harmonically
trapped dipolar condensates. Consider a pure dipolar condensate
polarized in the $\ds{z}$-direction. Note that due to the presence
of a Feshbach resonance, a practically pure dipolar system can be
produced by tuning the s-wave scattering length to zero. \\
A variational method can be employed to obtain an estimate for the
instability conditions. We will use the notation used in
\cite{gen8}.
\begin{equation}\label{notation}
\gamma{\rightarrow}\frac{1}{l^2}~,~~~\kappa{\rightarrow}\frac{1}{\kappa}~.
\end{equation}
We employ the following trial function
\begin{equation}\label{cm1}
\Psi(\mathbf{r})=\sqrt{\frac{N}{\pi^{3/2} \sigma^3d_{\perp}^3
\kappa}}~\exp \left( -\frac{1}{2 \sigma^2d_{\perp}^2} ( r^2+
\frac{z^2}{\kappa^2})\right)~,
\end{equation}
where $\ds{d_{\perp}=\sqrt{\hbar/m\omega_{\perp}}}$ is the
oscillator strength and the variational parameters $\ds{ \sigma}$
measures the change in radial dimension and $\ds{\kappa}$ is the
condensate aspect ratio.\\
The total energy $\ds{E^{0}=E^{0}_{dd}+E^{0}_{nd}}$, where
$\ds{E^{0}_{dd},E^{0}_{nd}}$ are dipolar and non-dipolar parts, is
given by
\begin{eqnarray}\label{2jbirlest}
\frac{E^{0}_{dd}}{E_{\perp}} &=& - \sqrt{\frac{2}{\pi}}\frac{N^{\star}}{\sigma^{3}}\left(\frac{1}{3\kappa}-\frac{ F(\kappa)-1}{\kappa(\kappa^2-1)}\right)~,\nonumber\\
\frac{E^{0}_{nd}}{E_{\perp}} &=&  (1+\frac{\kappa^2}{2
l^4}~)\sigma^{2}+(1+\frac{1}{2 \kappa^2}~) \sigma^{-2}~.
\end{eqnarray}
where $\ds{E_{\perp} =\frac{1}{2}N \hbar \omega_{\perp}}$, the
function
$\ds{F(\kappa)=\frac{\kappa\cosh^{-1}(\kappa)}{\sqrt{\kappa^2-1}}}$
is a continuous function ($\ds{\kappa>0}$) and
\begin{eqnarray}\label{nstar}
N^{\star}=\frac{m~ C_{dd}}{4\pi\hbar^2d_{\perp}}N~.
\end{eqnarray}
We investigate the existence of minimum of the total energy as a
function of the two variational parameters, $\ds{\sigma}$ and
$\ds{\kappa}$. The minimum conditions are given by
$\ds{\frac{\partial E}{\partial \sigma}=0}$ and
$\ds{\frac{\partial E}{\partial \kappa}=0}$. If we substitute the
total energy, $\ds{E^0}$, into these equations, we obtain
\begin{equation}\label{4685l8}
{N^{\star}}^4= \frac{\frac{324 \pi^2  (\kappa^2-1)^8l^{4}
}{(\kappa^4-l^4)^{-4}\kappa^{6}}\left(
(\frac{9}{\kappa^6}+\frac{18}{\kappa^2}) F-2-\frac{18}{\kappa^2}
-\frac{3}{\kappa^4}-\frac{4}{\kappa^6}\right)}{\left((\frac{9}
{\kappa^4}+\frac{18l^4}{\kappa^4}) F+
2-\frac{7}{\kappa^2}-\frac{4}{\kappa^{4}}-\frac{2l^4}{\kappa^2 }
(1+\frac{10}{\kappa^2}-\frac{2}{\kappa^{4}})\right)^5}
\end{equation}
\begin{equation}\label{e78no2}
\frac{\sigma}{N^{\star}}=\frac{(\frac{9}{\kappa^4}+\frac{2l^4}{\kappa^4}
)F+2-\frac{7}{\kappa^2}-\frac{4}{\kappa^4}-{\frac{2l^4}{\kappa^2}}
(1+\frac{10}{\kappa^2}-\frac{2}{\kappa^4}) }{3\sqrt{2\pi}~
\kappa(1-\frac{1}{\kappa^2})^2 (1-\frac{l^4}{\kappa^4})}
\end{equation}
The first one enables us to determine the condensate aspect ratio,
$\ds{\kappa}$, for a given number of particles and trap aspect
ratio. The second equation gives the information how the radial
and axial lengths change with $\ds{\kappa}$ and $\ds{l}$.\\
To find the critical number of particles, we should also calculate
the Hessian, $\ds{\Delta=\frac{\partial^2 E^0}{\partial
\sigma^2}\frac{\partial^2 E^0}{\partial
\kappa^2}-\left(\frac{\partial^2 E^0}{\partial \sigma
\partial \kappa}\right)^2}$. The local minimum of energy disappears and
instability of the condensate occurs when the Hessian is equal to
zero. The Hessian is too long to write here. Below, we will
predict the existence of a critical trap aspect ratio.

\subsection{Critical Points}

\begin{figure}[htp]
\includegraphics[width=6cm]{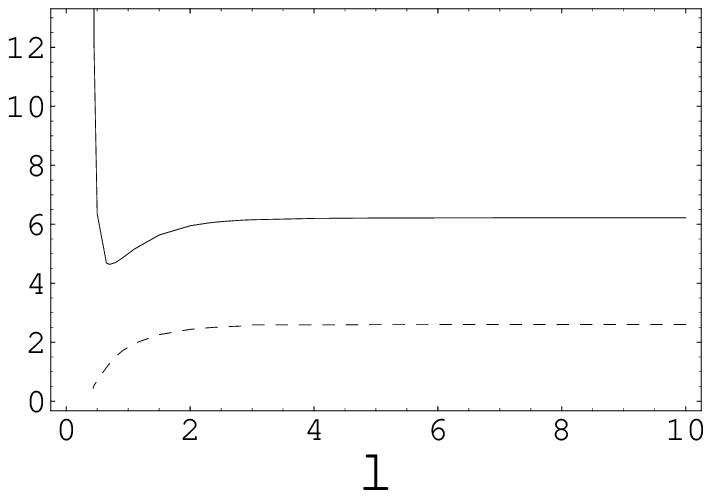}
\caption{\label{fig:sek0} The critical number of atoms,
$\ds{N_c^{\star}}$, (solid) and the critical value of the
condensate aspect ratio, $\ds{\kappa_c}$, (dotted) as a function
of the trap aspect ratio, $\ds{l}$, for a purely dipolar
condensate with $\ds{C_{dd}>0}$.}
\end{figure}

\begin{figure}[htp]
\includegraphics[width=6cm]{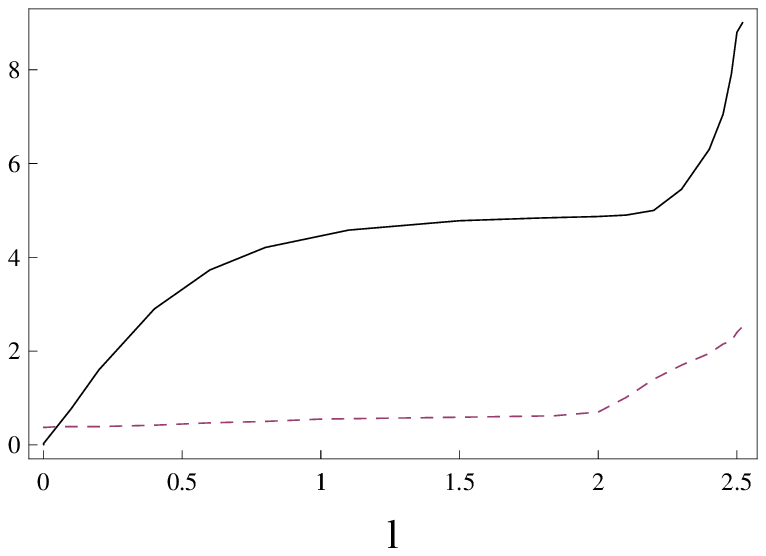}
\caption{\label{fig:CriticalNumber} The critical number of atoms,
$\ds{|N_c^{\star}|}$, (solid) and the critical value of the
condensate aspect ratio, $\ds{\kappa_c}$, (dotted) as a function
of the trap aspect ratio, $\ds{l}$, for a purely dipolar
condensate with $\ds{C_{dd}<0}$.}
\end{figure}

There exists two critical values for the trap aspect ratio
\begin{equation}\label{crit}
l^{\star}=0.43~,~~~~l^{\star \star}=2.52~.
\end{equation}
Only the first critical point in Gaussian limit was predicted in
the literature \cite{gen8}. The critical trap aspect ratios play
important roles in understanding of the general structure of
dipolar condensates. Firstly, the dipole-dipole interaction is
positive when $\ds{l<l^{\star }}$ ($\ds{l>l^{\star\star}}$) for
$\ds{C_{dd}>0}$ ($\ds{C_{dd}<0}$). Hence, the dipolar condensate
will be stable at any number of particles as long as
$\ds{l<l^{\star }}$ for $\ds{C_{dd}>0}$ and
$\ds{l>l^{\star\star}}$ for $\ds{C_{dd}<0}$. The dipole-dipole
interaction is negative when $\ds{l>l^{\star\star}}$
($\ds{l<l^{\star }}$) for $\ds{C_{dd}>0}$ ($\ds{C_{dd}<0}$). So,
the stability requires $\ds{N^{\star}<N_c^{\star}}$ where
$\ds{N_c^{\star}}$ is the critical a critical value. The critical
number $\ds{N_c^{\star}}$, contrary to the case of contact
interaction, depends strongly on the trap geometry. Let us suppose
$\ds{\kappa=\kappa_c}$ and $\ds{\sigma=\sigma_c}$ when
$\ds{N^{\star}=N_c^{\star}}$. It is interesting to observe that
$\ds{N_c^{\star}{\approx}6.2}$, $\ds{\kappa_c{\approx}2.6}$, and
$\ds{\sigma_c{\approx}0.68}$ remain almost the same when
$\ds{l>3}$ for $\ds{C_{dd}>0}$. Fig-\ref{fig:sek0} and
Fig-\ref{fig:CriticalNumber} plot $\ds{|N_c^{\star}|}$ (solid
curve) and $\ds{\kappa_c}$ (dashed curve) as a function of trap
aspect ratio for $\ds{C_{dd}>0}$ and $\ds{C_{dd}<0}$,
respectively. Amazingly, $\ds{N_c^{\star}}$ and $\ds{\kappa_c}$
change sharply at $\ds{l=0.8}$ and $\ds{l=0.7}$, respectively for
$\ds{C_{dd}>0}$ and at $\ds{l=2.3}$ and $\ds{l=2.1}$, respectively
for $\ds{C_{dd}<0}$. As can be seen from the figures, there exists
a minimum for critical number, $\ds{N_c^{\star}=4.64}$ at
$\ds{l=0.7}$ for $\ds{C_{dd}>0}$, while no such minimum occurs for
$\ds{C_{dd}<0}$. We note that in the region
$\ds{l^{\star}<l<l^{\star\star}}$, the sign of the dipole-dipole
interaction  depends on $\ds{C_{dd}}$ and $\ds{l}$.\\
Secondly, the condensate aspect ratio is equal to the trap aspect
ratio, $\ds{\kappa=l}$, only if $\ds{l=l^{\star }}$ or
$\ds{l=l^{\star \star}}$. Only in the region
$\ds{l^{\star}<l<l^{\star\star}}$, $\ds{\kappa}$ is bigger
(smaller) than $\ds{l}$ for $\ds{C_{dd}>0}$ ($\ds{C_{dd}<0}$). As
$\ds{N^{\star}}$ is increased, $\ds{\kappa}$ increases (decreases)
when $\ds{l^{\star}<l<l^{\star\star}}$ ($\ds{l<l^{\star }}$ or
$\ds{l>l^{\star\star }}$) for $\ds{C_{dd}>0}$. Conversely, as
$\ds{N^{\star}}$ is increased, $\ds{\kappa}$ decreases (increases)
when $\ds{l^{\star}<l<l^{\star\star}}$ ($\ds{l<l^{\star }}$ or
$\ds{l>l^{\star\star }}$) for $\ds{C_{dd}<0}$. The dipole-dipole
interaction stretches (squeezes) the cloud both in the radial and
the axial directions compared to the non-dipolar condensate if
$\ds{l<l^{\star }}$ for $\ds{C_{dd}>0}$ ($\ds{C_{dd}<0}$) and if
$\ds{l>l^{\star\star }}$ for $\ds{C_{dd}<0}$ ($\ds{C_{dd}>0}$).\\
It is remarkable to note that the existence of two critical trap
aspect ratios, $0.44$ and $2.30$ (in our notation), for a
metastable TF solution outside of the regime
$\ds{-0.5<\varepsilon_{dd}<1}$ was discussed \cite{duzeltme}. It
is interesting to observe that these critical values in TF
approximation are very close $\ds{l^{\star}}$ and
$\ds{l^{\star\star}}$ in Gaussian approximation (\ref{crit}). It
was shown that collapse via scaling oscillations is suppressed in
TF limit if $\ds{l>2.30}$ for
$\ds{\varepsilon_{dd}\rightarrow{-\infty}}$ and if $\ds{l<0.44}$
for
$\ds{\varepsilon_{dd}\rightarrow{\infty}}$ \cite{duzeltme}.\\
In the following two subsections, we will show that the critical
trap aspect ratios are changed by an additional anharmonic
potential
\begin{equation}\label{vrd}
V_{T}=\frac{m}{2}\omega_{\perp}^2\left(r^2+ ~\frac{z^2}{l^4}
+\lambda_{\perp}~ \frac{r^4}{d_{\perp}^2}+\lambda_z~
\frac{z^4}{l^6 d_{\perp}^2}\right)~,
\end{equation}
where $\ds{\lambda_{\perp}}$ and $\ds{\lambda_{z}}$ are small
dimensionless parameters characterizing the strength of the
quartic potential in radial and axial directions, respectively.
This potential has been achieved experimentally by superimposing a
blue detuned laser beam  with the Gaussian profile to the magnetic
trap \cite{deney}.

\subsection{Axially Quartic Potential}

We will now investigate the critical trap aspect ratios in the
presence of anharmonic potential. Suppose first that anharmonic
potential is in the axial direction, $\ds{\lambda_{\perp}=0}$. The
selection of the proper form of trial functions is very important
in the variational approach. We will choose a trial function which
reduces to the ground state solution of the Schrodinger equation
in the noninteracting limit. Hence, we employ the trial function
up to the first order of $\ds{\lambda_{z}}$
\begin{equation}\label{c3211h}
\Psi=\psi_0~\exp \left( \frac{-1}{2 \sigma^2d_{\perp}^2} ( r^2+
\frac{z^2}{\kappa^2}+\frac{\lambda_z}{4 \kappa^2 }
(3z^2+\frac{z^4}{\kappa^2\sigma^2}))\right)
\end{equation}
where $\ds{\psi_0=\sqrt{\frac{N}{\pi^{3/2} \sigma^3d_{\perp}^3
\kappa}}~(1+\frac{9}{32}\lambda_z)}$ is the normalization
constant. If we evaluate the energy expression, $E^{\lambda_{z}}=
E_{nd}^{\lambda_{z}} +E_{dd}^{\lambda_{z}}$, we get
\begin{eqnarray}\label{enerbirlest}
\frac{E_{dd}^{\lambda_{z}}}{E_{\perp}}&=&\frac{E^{0}_{dd}}{E_{\perp}}-\sqrt{\frac{2}{\pi}}
N^{\star} \lambda_z \frac{G_{z}(\kappa)
+3\kappa^2(24-19\kappa^2)F(\kappa)}{32\kappa(\kappa^2-1)^3}\nonumber\\
\frac{E_{nd}^{\lambda_{z}}}{E_{\perp}}&=&
\frac{E^{0}_{nd}}{E_{\perp}}+\frac{3\lambda_z}{4}\left(\frac{1}{\kappa^{2}\sigma^{2}}+\frac{\sigma^{2}
\kappa^2}{l^{6}}(\sigma^{2} \kappa^2-l^2)\right)
\end{eqnarray}
where  $\ds{E_{\perp} =\frac{1}{2}N \hbar \omega_{\perp}}$,
$\ds{E^{0}_{dd}}$ and $\ds{E^{0}_{nd}}$ are the dipole-dipole and
non-dipole energies for purely harmonic trap (\ref{2jbirlest}) and
$\ds{G_{z}(\kappa)=14-88\kappa^2+52\kappa^4+7\kappa^6}$. The
minimums of the variational parameters $\ds{\sigma}$ and
$\ds{\kappa}$ can be calculated from these energy expressions.
Then critical aspect ratios can be found. We find that they are
decreased with $\ds{\lambda_z}$. For example,
\begin{equation}\label{gzdydfyd}
\begin{array}{ccccc}
l_{\lambda_z=0}^{\star}=0.43; & ~l_{\lambda_z=0.05}^{\star}=0.41;  & ~l_{\lambda_z=0.1}^{\star}=0.39.  \\
\end{array}
\end{equation}
\begin{equation}\label{54fdufb5}
l_{\lambda_z=0}^{\star \star}=2.52,~~ l_{\lambda_z=0.05}^{\star
\star}=2.43 ~~ l_{\lambda_z=0.1}^{\star \star}=2.34~.
\end{equation}
The critical number of particles, $\ds{N^{\star}_{c}}$ doesn't
change appreciably with $\ds{\lambda_z}$. When $\ds{l>l^{\star
\star}}$, the critical values of the condensate aspect ratio,
$\ds{\kappa_{c}}$, is decreased with $\ds{\lambda_z}$ for
$C_{dd}>0$. For example, if ${l=10}$ then
$\kappa_{c,\lambda_z=0}=2.60$, $\kappa_{c,\lambda_z=0.05}=2.49$
and $\kappa_{c,\lambda_z=0.1}=2.40$. However, $\sigma_c\approx
0.68$ slightly increases with $\ds{\lambda_z}$.

\subsection{Radially Quartic Potential}

Suppose now that anharmonic potential is in the radial direction,
$\ds{\lambda_{z}=0}$. We will choose our trial function up to the
fist order of $\ds{\lambda_{\perp}}$
\begin{equation}\label{c3211hlo}
\Psi= \psi_0\exp \left( \frac{-1}{2 \sigma^2d_{\perp}^2} ( r^2+
\frac{z^2}{\kappa^2}+\lambda_{\perp}
(r^2+\frac{r^4}{4\sigma^2d_{\perp}^2}))\right)
\end{equation}
where
$\ds{\psi_0=\sqrt{\frac{N}{\pi^{3/2}d_{\perp}^3\sigma^3\kappa } }
(1+\frac{3}{4}\lambda_{\perp}) }$ is the normalization constant.
The corresponding energy, $E^{\lambda_{\perp}}=
E_{nd}^{\lambda_{\perp}} +E_{dd}^{\lambda_{\perp}}$, is given by
\begin{eqnarray}\label{enerbirlest}
\frac{E_{dd}^{\lambda_{\perp}}}{E_{\perp}}&=&
\frac{E_{dd}^0}{E_{\perp}}-\sqrt{\frac{2}{\pi}} N^{\star}
\lambda_{\perp} \frac{G_{\perp}(\kappa)
+9(32\kappa^2-27)F(\kappa)}{96\kappa(\kappa^2-1)^3}~,\nonumber\\
\frac{E_{nd}^{\lambda_{\perp}}}{E_{\perp}}&=&
\frac{E_{dd}^0}{E_{\perp}}
+2\lambda_{\perp}\left(\sigma^{4}-\sigma^{2}+\sigma^{-2}\right)~,
\end{eqnarray}
where
$\ds{G_{\perp}(\kappa)=112+45\kappa^2-258\kappa^4+56\kappa^6}$.\\
From the energy expression, we can find the critical trap aspect
ratios. The first critical point changes very slowly with
$\ds{\lambda_{\perp}}$ in contrast to the case of axially
anharmonic potential (\ref{gzdydfyd}).
\begin{equation}\label{9ueyurp}
l^{\star}_{\lambda_{\perp}=0}=0.43~,~~l^{\star}_{\lambda_{\perp}=0.1}=0.44~.
\end{equation}
The second critical point is increased with
$\ds{\lambda_{\perp}}$. For example,
\begin{equation}\label{54fdu}
l_{\lambda_{\perp}=0}^{\star \star}=2.52,~
l_{\lambda_{\perp}=0.05}^{\star \star}=2.56 ~
l_{\lambda_{\perp}=0.1}^{\star \star}=2.61~.
\end{equation}
$\ds{N_{c}}$ and $\ds{\sigma_{c}}$ doesn't change appreciably with
$\ds{\lambda_{\perp}}$. When $\ds{l>l^{\star \star}}$,
$\ds{\kappa_c}$ is increased due to the quartic potential. As an
example, if $\ds{l=10}$, then
$\ds{\kappa_{c,\lambda_{\perp}=0}=2.6}$,
$\ds{\kappa_{c,\lambda_{\perp}=0.05}=2.7}$ and
$\ds{\kappa_{c,\lambda_{\perp}=0.1}=2.8}$.\\
In this paper, we have shown that there exists some critical trap
aspect ratios for a dipolar condensate determining the general
structure of the condensate. Classifying the trap as prolate and
oblate is not helpful. It was generally believed that critical
angular velocity, $\ds{\Omega_c}$, decreases (increases) due to
dipolar interaction in an oblate (a prolate) trap. Here, we have
discussed that there exists a critical trap aspect ratio
${\gamma_c{\approx}2.8}$ above (below) which $\ds{\Omega_c}$ is
increased (decreased) due to dipolar interaction. We have also
shown that the same is true for the precession velocity of an
off-axis vortex. Furthermore, we have proved that there are two
critical trap aspect ratios for a purely dipolar gas in Gaussian
limit. The stability diagram depends crucially on the critical
trap ratios. We have shown that the critical points can be changed
in the presence of quartic potential.\\
Helpful discussions with Z. Oztas are gratefully acknowledged.

\end{document}